# Secure Hybrid ITS Communication with Data Protection

**Paper ID EU-TP0875**

## Secure Hybrid ITS Communication with Data Protection

**Jonas Vogt[1], Manuel Fünfrocken[1], Niclas Wolniak[1], Prof. Dr. Horst Wieker[1*]**
1.  htw saar, University of Applied Sciences, Saarbrücken, Germany, Goebenstr. 40, 66117 Saarbrücken, +49 681 5867 195, wieker@htwsaar.de

**Abstract**
In the future world of safe traffic, intelligent transportation systems play a vital role. The connections between traffic participants and systems on the one and infrastructure and service providers on the other side are a necessary prerequisite for informed and safe driving. The key for the connection between all stakeholders is a reliable and secure connection. Reliability can be achieved in different ways. For the information exchange, we utilize hybrid communication technologies in different scenarios. This includes long-range technologies like DAB+ and cellular networks and short- and medium-range technologies like ETSI ITS G5 (IEEE 802.11p) or RFID. Those technologies can only be used if messages transmitted via the wireless communication links are protected. Protection not only means encryption but most importantly means to verify if the data was sent by a legit sender. This is a necessary requirement for a recipient to trust the data received. In ITS, transmitted data could be used for user tracing and collection of user data. Therefore, a communication system should be designed in such a way that personal user data is protected. The user must be untraceable, but nevertheless able to use all services without limitations. In this paper, we describe the approach and ideas we took to design a secure hybrid communication architecture that is at the same time data protection friendly.

**Introduction**
Highly automated driving functions can enable more efficient electric driving and increase the safety of the vehicle occupants. However, now there are still inhibitions, which lead to market barriers for automated functions in electric vehicles. In contrast to the "classic" mobility with combustion engines, electric mobility addresses a significantly wider environment. The smart interplay of intelligent traffic systems in the infrastructure (Smart Traffic), the intelligent and automated systems of an e-vehicle (Smart eCar) and the energy infrastructure (Smart Grid) is key for electric mobility. We focus on the electric vehicle.
The currently available systems in the market do not pursue a comprehensive and consistent approach. The fields of electric mobility and highly automated driving currently offer neither transparent communication structures nor consistent quality of information. Functions cannot be transferred manufacturer-independently and technologies are not coordinated with one another. The integration of third-party service providers (such as car park operators) as well as infrastructure facilities (such as traffic lights, loading infrastructure, etc.) is only possible through proprietary approaches. So far, the efficient use of broadcast-based systems for the dissemination of information in the field of electric mobility has not been considered at all. This makes the sensible integration of manufacturer-wide automatic functions for efficient electric driving more difficult.
The work in this paper was done in the context of the German research project iKoPA [1]. The aim of iKoPA is to combine different infrastructure systems. The basis for this is a multi-core communication approach with the communication standards ETSI ITS-G5 (IEEE 802.11p), DAB+ TPEG and cellular communication. Therefore, we aim to dissolve the existing proprietary structures and to integrate the various technologies. Uniform access points for service providers and access networks are specified. Secure access to the

## Secure Hybrid ITS Communication with Data Protection

network is ensured without any problems. With regard to the global goal of achieving efficient, safe and highly automated electric driving, the communication, service and organizational architecture for the use in the electric mobility environment has to be adapted and expanded. The work is based on the results from the CONVERGE research project. iKoPA follows an innovative, multi-faceted communication approach with simultaneous support for vehicle-to-x communication, DAB TPG and mobile communications and the integration of driver assistance system architectures to support highly-automated driving manoeuvres.
On top of this, we will also address the integration of security and data protection, which will be described below. To challenge real conditions, the iKoPA architecture will be implemented in a demonstration scenario.

### Technical Goals and Communication

Communication technologies might be suitable to reduce problems of electric mobility, thus increasing its attractiveness to consumers. When talking about communication, a couple of needed basic characteristics, or goals, come to mind. The first one, easily understandable when considering recent `accomplishments` in the field of remotely exploiting vulnerabilities in vehicle architectures [2], is security. Each communication based solution, which interacts with the traffic system and vehicles in particular needs to have comprehensive security in place. As the technology inventions in the `internet age` have shown, it is not easy, and often not possible, to add reliable, strong security to an existing system. Therefore, security considerations need to be taken into account on all levels of the architectures, as well as from an overarching security viewpoint, to make sure end-to-end security in the final system is possible. Closely related to security is the topic of data protection, which will be more explicitly described in the section below. According to the dominant scientific opinion in information and communication technologies, it is not possible to foresee the prevalent technology of the coming decade today. For sure, there will be innovations and adaptions, changing current technologies and introducing new ones. With the comparable long clock-speed of the automotive industry and the long usage times of vehicles, a sustainable architecture needs to be able to adapt to changing technologies. In previous research projects [3,4] it has been concluded that a hybrid communication approach, which combines different communication technologies, is necessary. This conclusion was also considered the European Commission in their strategy white paper for C-ITS in Europe. [5] Designing architectures with changing and complementary technologies in mind from the beginning ensures that they can be adapted in the future and that it becomes possible to combine different technologies to compensate the weaknesses of one technology with the strengths of another.
In iKoPA, three different communication technologies are used: cellular communication using ETSI IMT Public (3G, "LTE"), vehicle-to-infrastructure communication using ETSI ITS G5 ("V2X") and broadcast via (terrestrial) Digital Audio Broadcast (DAB+). The combination of the first two technologies provides bidirectional communication to traffic participants in a multitude of situations. With the integration of V2X communication, information to enhance to situational awareness of vehicles can be sent in situations, where infrastructure bound communication is not possible. For example, one use case for this is the provision of enhanced positioning information to automatic vehicles inside of parking garages. In this environment, positioning is still a relevant task, whereas traditional technologies like cellular communication or access to satellite navigation systems is usually unreliable or entirely impossible.
By integrating DAB+, a broadcast medium to efficiently reach a huge number of traffic participants is added to the system. Not only provides DAB+ one of the most efficient ways of communication regional traffic information to a multitude of receivers, its usage does also ensure, that the mechanisms of the architecture are able to cope with broadcast networks in general.
By providing all those communication mechanisms, it gets easy for electric vehicles to communicate with the infrastructure, e.g. for reserving charging points or planning a route. Moreover, automatic, electric vehicles can get access to a more elaborate information base to make strategic driving decisions.

### Data Protection

In the current world of ITS and ICT the amount of data collected and stored is increasing every day. Many approaches for newly developed solutions for automated driving and intelligent traffic are based on crowd and big-data solutions (e.g. BMW Connected Drive, GM OnStar or Here Maps). In such applications, a huge amount of data from every participant is sent to a cloud service for processing and storage. The data transmitted is often pseudonymized but can qualify as personal data [5] of the user/vehicle generating this data. In addition, the number of data sets and the purpose for which the data is collected is not always clear. This leads to a collision between the allegedly necessary data for such applications and the

# Secure Hybrid ITS Communication with Data Protection

fundamental rights of the user. In Germany, the right to informational self-determination is the fundamental right of every person to determine the disclosure and use of his or her personal data. In the European Union personal data is protected by Article 8 of the 'Charter of Fundamental Rights of the European Union' [6]. In general, the right for privacy is fundamental motivated also on §12 of the United Nations 'Universal Declaration of Human Rights' [7].

When building an ITS architecture the protection of all users´ personal data should be a basic keystone. To support the development six protection goals [8,9] are defined. These are the three IT-security goals

- confidentiality (requirement that no person can learn about personal data unauthorized),
- integrity (requirement that information technology processes and systems comply with the specifications that have been defined for the purpose of performing their intended functions for them) and
- availability (requirement that personal data must be available and can be used properly in the intended process)

and the three data protection specific goals:

- unlinkability (requirement that data be processed and evaluated only for the purpose for which it is collected),
- transparency (requirement that both affected persons, as well as the operators of systems as well as competent supervisory authorities can validate, which data for what purpose in a procedure is collected and processed, and which systems and processes are used) and
- intervenability (requirement that the data subjects (aka the users) are effectively granted the rights of notification, information, correction, blocking and deletion which they are entitled to at all times, and that the processing body is obliged to implement the appropriate measures).

These six goals are accompanied by the principle of data minimization.

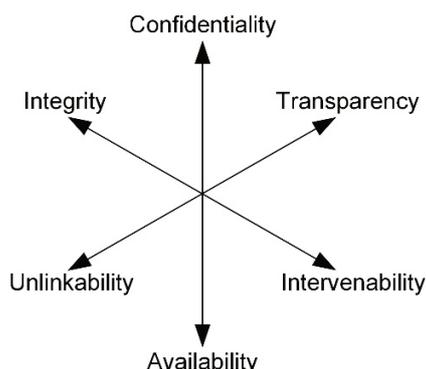

**Figure 1: Six protection goals for privacy**

Following the data protection path numerous requirements have to be considered when designing an ITS communication architecture. In the following, we will only focus on key requirements. First, the data controllers involved need a legal ground for data processing and data collecting. The data involved has to be specified and the purpose of the collection and processing must be evident and is not allowed to change without a legal ground or the consent of the data subject. The available legal grounds have the principle of data minimization in common. This means, that only those data sets are gathered that are necessary for a service. For the data collected and stored by the service provider, the user has to be able to review, change and delete all data or parts of it. In addition, the user has to be able to transfer its data from one service provider to another service provider ('right to data portability'). This transfer should be possible in an electronic manner and basic user settings should be preserved during the transfer. For this purpose, an architecture should foresee an entity that serves as a single point of contact for a user regarding data protection topics. This entity should also serve as a mediator between the parties involved.

Normally, a user is not only using one service, but a number of different services. To protect the privacy of the user, the usage of one service should not be linkable to the usage of another service. For this purpose, a pseudonym service usage is required. The service provider should not need to know who the user is but only that it is a legit user.

Independent from the service and the user, this leads to some general architecture requirements. First, an

## Secure Hybrid ITS Communication with Data Protection

architecture should avoid central entities, were huge amount of data is stored or processed and could be combined. Additionally, a single central entity also diminishes the availability of the system. The data stored has to be encrypted and the access to this data must be minimized. The data must be verified and any change should be documented. For communication, it is also required to encrypt and sign the message, so that no third party can view or change the data during transfer.

In conclusion, the system should be designed by the principle 'privacy by design'. This must be considered during the architecture phase for the communication architecture and the services using that architecture, but also and especially during the operational phase of the services. Finally, all processes and procedures have to be documented and available for inspection at any time.

**IT-Security**

The success of modern communication systems relies, besides the designed communication links, on the security that can be ensured on this links but also in the systems connected by the communication architecture. Therefore, security has to be included in the design of the architecture by day one. At the same time, security also has the means necessary to include privacy techniques into the communication architecture. Our security focus has two parts:

- Secure and privacy friendly communication
- System security

The communication is modelled on several layers in the different parts of the architecture. From a communication perspective, the architecture is roughly parted in three planes (see next chapter for more details):

- the backend/governance plane (with all the service providers and management facilities),
- the communication network plane (including the communication networks itself, but also service related components that are included in those networks), and
- the remote station plane (vehicles, smartphones, but also charging stations and other (traffic) infrastructure components).

For the communication inside of and between the different planes, several communication technologies are used in the context of the project. At the backend plane, the cable-based connections of the internet are used. Through the communication networks and especially to and inside the remote plane the following technologies are included:

- cellular (for mostly single connections between service users and service providers, used for information specific for one user or relevant in a big area),
- ETSI ITS G5 (ad-hoc communication up to a few hundred meters, used for time critic and local relevant information),
- DAB+ (digital audio broadcast, used for information from the service providers to the service users for information that are intended for a large amount of users in a big area; no back channel), and
- RFID (short range radio-frequency identification up to a few meters, used for authentication of users).

The goal of the security architecture is to facilitate and if necessary to extend the existing security mechanism to describe a consistent approach over all communication technologies while incorporating the similarities and differences. This approach is especially challenging because of the aspiration to make the connections privacy friendly and protect the privacy of the user on all levels. Therefore, different existing security and privacy techniques have to be combined.

To ensure the security and data protection not only the connection between systems but also the systems itself need to be considered. Cryptographic algorithms have to be implemented correctly, keys have to be stored securely, systems have to be protected against attacks, and user data have to be handled and secured according to data protection law. Whereas the protection of the user data from the point of privacy is not only a technical but also an organizational matter, the secure storage of keys and the crypto algorithm are essential technical parts. For that reason, a secure platform based on a TPM 2.0 (trusted platform module) will be developed. This module should be designed in such a way that it can be integrated on all planes in all systems that participate in communication or hold personal user data. Meaning it can equivalently be integrated in all systems from smartphone, vehicles and charging stations to communication network equipment and backend servers. Additionally, a consistent API (application programming interface) should be provided, so that services can use the interface independent of the system they are running.

# Secure Hybrid ITS Communication with Data Protection

## Architecture Approach

In the iKoPA project the CONVERGE architecture is taken as a basis, as shown in Figure 2. The goal is to integrate the technical requirements into the existing architecture in terms of privacy, IT security and e-mobility as described in the previous sections. The architecture defines the following characteristics:

- **Open**: accessible for all systems, which fulfill a specific set of criteria, including a certification process for new participants and new services.
- **Distributed:** The components of the Car2X Systems Network spread across several participants and are placed in different locations, so that they do not depend on each other.
- **Trans-regional connected:** As vehicles travel between different regions and countries, an ITS system cannot be restraint to one region, but has to be usable in many regions and countries. Therefore, it must support information exchange between the systems in different regions and across political and organizational borders.
- **Provider independent:** The functionality of the network should not depend on a single provider.
- **Scalable:** The network must be able to handle a handful of services and a few hundred participants in the starting phase as well as thousands of services and maybe millions of users in the later years.
- **Hybrid communication**: The network must work with different communication technologies to connect service users and providers.
- **Flexible:** As most of the services are not known in advance, there must not only be a technological flexibility as described above, but also an economical flexibility, which encourages service providers to develop and introduce new services and business cases.
- **Secure:** The network should ensure the safety, security and the privacy of the users, the service providers and the communication between them.

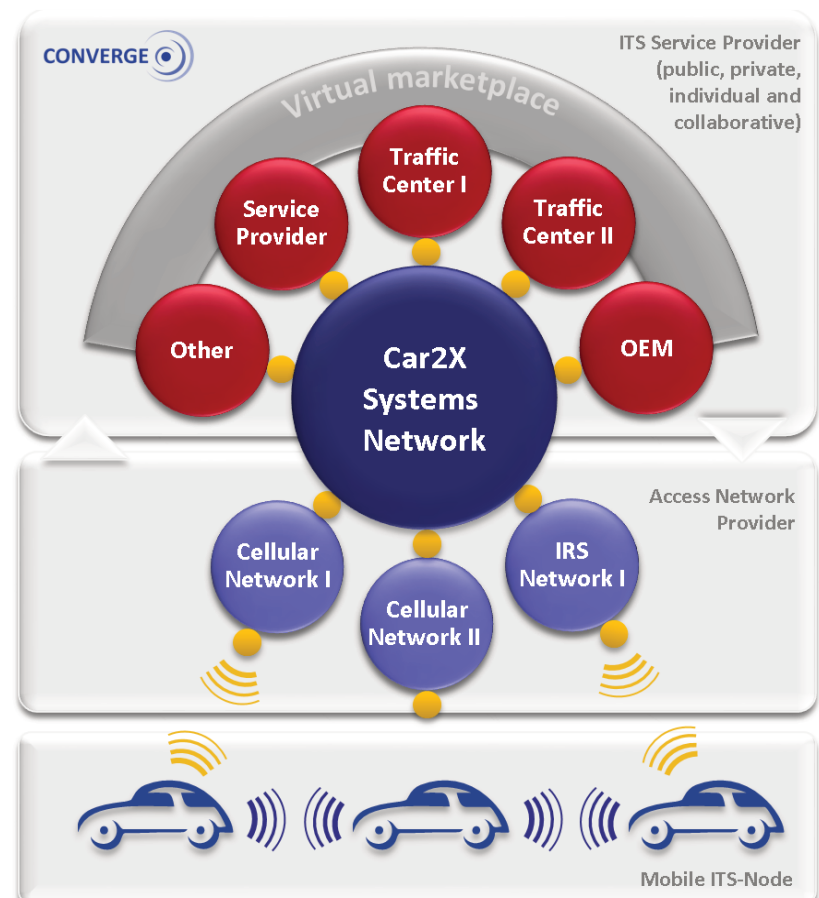

**Figure 2: CONVERGE architecture**

# Secure Hybrid ITS Communication with Data Protection

Focusing on the so called 'Car2X Systems Network', the architecture has the goal to break up the current fixed connections between access networks and service providers by specifying standardized, uniform access points between communication networks and service providers.

In extension to the CONVERGE architecture, we strongly focusses on privacy, IT security and the integration of existing communication technologies into a hybrid communication architecture. From a communication perspective, we will introduce DAB+ and RFID to the architecture, which currently covers mobile communication and ETSI ITS G5 (IEEE 802.11p). Furthermore, the architecture (see Figure 3) will be focusing on integrating infrastructure components, which are not part of a communication network, alongside with mobile nodes into the architecture, e.g. a charging station or an access control barrier. Therefore, defining the lower layer of the architecture as the "Remote Node" layer in contrast to the "Mobile ITS-Node" layer in the CONVERGE architecture, which was rather seen as a given condition for the architecture in the CONVERGE project. Therefore, remote nodes define end-points of a communication connection at a physical field level.

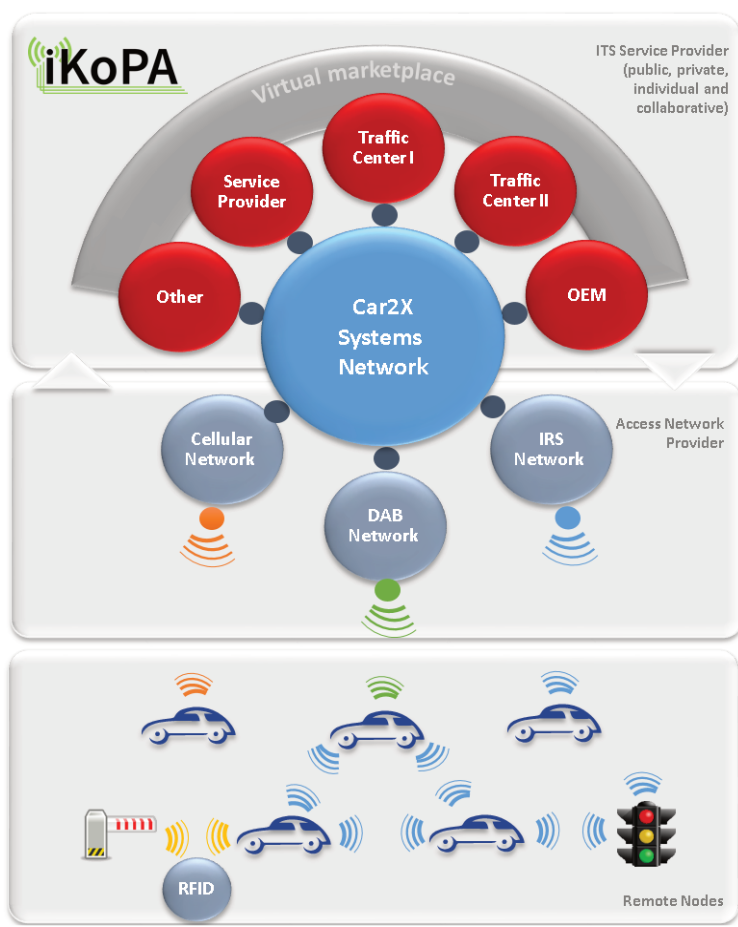

**Figure 3: iKoPA architecture**

Summarized, the architecture consists of three planes. The backend plane on the top with all its services providers, infrastructure systems, and security and management components. In the middle is the communication plane with the access network providers. Currently three access networks (cellular, DAB+, ITS-G5 via IRS) are included but further networks can be easily integrated at a later point. The third plane on the bottom are the remote nodes. These are e.g. vehicles, smartphones, access barriers, charging stations and traffic light. The systems in the remote plane interchange information with the communication plane but also with systems on the same plane.

# Secure Hybrid ITS Communication with Data Protection

**Impact**
The main result is the communication architecture itself, made available in the form of multiple reports, and shown and evaluated during a demonstration. This architecture will more precisely than previous architectures describe how the various parties need to interact with each other. Of key interest in this architecture are the considerations to privacy, integration of DAB+ and the possibilities for electric mobility. With an ever-growing urge by companies and policy makers to collect and evaluate more and more personal data, the privacy protecting mechanisms of the proposed architecture will provide solutions to implement mechanisms, which consider data minimization. Together with the security mechanisms this ensures, that the privacy of future users is not jeopardized by the system.
With the integration of DAB+, a broadcast network is integrated into the C-ITS architecture. This provides a standardized mechanism to easily reach many participants with a minimal communication overhead. Through the interconnection of charging infrastructure, traffic infrastructure and vehicles, applications for smarter and greener mobility can be developed more easily in the future.

**Facts and acknowledgment**
The iKoPA project will run from 2015-12-01 until 2018-11-30. The consortium consists of Bayerische Medien Technik GmbH, Daimler Center for Automotive Information Technology Innovations, Fraunhofer Institute for Open Communiction Systems FOKUS, Fraunhofer Institute for Secure Information Technology SIT, htw saar – Hochschule für Technik und Wirtschaft des Saarlandes - University of Applied Sciences (project lead), NXP Semiconductors Germany GmbH, SWARCO Traffic Systems GmbH, Unabhängiges Landeszentrum für Datenschutz Schleswig-Holstein.
This work was funded within the project iKoPA by the German Federal Ministry of Education and Research. The results presented in this paper were developed jointly by the iKoPA project partners.